\documentclass[12pt,preprint]{aastex}

\usepackage{graphics}

\def\gtrsim{\mathrel{\hbox{\rlap{\hbox{\lower4pt\hbox{$\sim$}}}\hbox{\raise2pt\hbox{$>$}}}}}

\newcommand{\mbh}{\ensuremath{M_\mathrm{BH}}}

\newcommand{\msun}{\ensuremath{M_{\odot}}}

\slugcomment{Accepted by The Astrophysical Journal}

\shorttitle{AGNs in low-mass galaxies}
\shortauthors{Schramm et al.}

\begin{document}


\title{Unveiling a population of galaxies harboring low-mass black holes with X-rays}


\author{M. Schramm\altaffilmark{1}, J. D. Silverman \altaffilmark{1}, J. E. Greene\altaffilmark{2}, W. N. Brandt\altaffilmark{6,7},B. Luo\altaffilmark{6,7}, Y. Q. Xue\altaffilmark{8}, P. Capak\altaffilmark{4}, Y. Kakazu\altaffilmark{5}, J. Kartaltepe\altaffilmark{9}, V. Mainieri \altaffilmark{3}}

\email{malte.schramm@ipmu.jp}

\altaffiltext{1}{Kavli Institute for the Physics and Mathematics of the Universe (WPI), Todai Institutes for Advanced Study, the University of Tokyo, Kashiwanoha 5-1-5, Kashiwa-shi, Chiba 277-8568, Japan}
\altaffiltext{2}{Alfred P. Sloan Fellow, Department of Astrophysical Science, Princeton University, Princeton, NJ, 08544, USA}
\altaffiltext{3}{European Southern Observatory, Karl-Schwarzschild-Strasse 2, Garching, D-85748, Germany}
\altaffiltext{4}{Spitzer Science Center, 314-6 California Institute of Technology, 1201 E. California Blvd., Pasadena, CA, 91125} 
\altaffiltext{5}{Department of Astronomy, California Institute of Technology, Pasadena, CA 91125, USA}
\altaffiltext{6}{Department of Astronomy and Astrophysics, Pennsylvania State University, University Park, PA 16802, USA }
\altaffiltext{7}{Institute for Gravitation and the Cosmos, Pennsylvania State University, University Park, PA 16802, USA }
\altaffiltext{8}{Key Laboratory for Research in Galaxies and Cosmology, Department of Astronomy, University of Science and Technology of China, Chinese Academy of Sciences, Hefei, Anhui 230026, China} 
\altaffiltext{9}{National Optical Astronomy Observatory, 950 North Cherry Avenue, Tucson, AZ 85719, USA }


\begin{abstract}

We report the discovery of three low-mass black hole candidates residing in the centers of low-mass galaxies at $z<0.3$ in the $Chandra$ Deep Field - South Survey.  
These black holes are initially identified as candidate active galactic nuclei based on their X-ray emission in deep Chandra observations. Multi-wavelength observations are used to strengthen our claim that such emission is powered by an accreting supermassive black hole.
While the X-ray luminosities are low at $L_X\sim10^{40}$ erg s$^{-1}$ (and variable in one case), we argue that they are unlikely to be attributed to star formation based on H$\alpha$ or UV-fluxes.  
Optical spectroscopy with Keck/DEIMOS and VLT/FORS allows us to (1) measure accurate redshifts, 
(2) confirm their low stellar host mass, (3) investigate the source(s) of photo-ionization, and (4) estimate extinction. 
With stellar masses of $M_*<3\times10^{9}$ M$_{\sun}$ determined from HST/ACS imaging, the host galaxies are among the lowest mass systems known to host actively accreting black holes.  
We estimate BH masses $M_{\rm BH}\sim 2\times 10^{5}$ M$_{\sun}$ based on scaling relations between BH mass and host properties for more luminous systems.  
In one case, a broad component of the H$\alpha$ emission-line profile is detected thus providing a virial mass estimate.   
Black holes in such low-mass galaxies are of considerable interest as the low-redshift analogs to the seeds of the most massive BHs at high redshift which have remained largely elusive to date.  
Our study highlights the power of deep X-ray surveys to uncover such low-mass systems.

\end{abstract}





\section{Introduction}

At the present time, nearly all massive galaxies contain 
supermassive ($10^6-10^9~\msun$) black holes (BHs) at their center.  The
formation mechanism of the seeds for these BHs is not known. They may have formed
from the death of the first, massive stars \citep[e.g.,][]{li2007,ohkubo2009,brommyoshida2011}, from the direct collapse of gas clouds
\citep[e.g.,][]{Koushiappasetal2004}, or in the centers of dense
stellar clusters \citep[e.g.,][]{devecchi2009}.  Whatever the
mechanism, we know that in some cases the seed BHs were able to gain
mass quickly to form massive quasars at $z\sim 6$
\citep[e.g.][]{fanetal2001}.  Additional insight into the formation of the
first BHs comes from looking for the least massive BHs in the Universe
today.  These are, in some sense, ''left-over'' seed BHs.  A study of
their number density, mass distribution, and host-galaxy properties
can in principle isolate the primary seed formation mechanism
\citep[e.g.,][]{volonterinatarajan2009,vanWass2012}.

Unfortunately, low-mass BHs (\mbh$ <10^6$~\msun) are difficult to study since they tend to live in faint galaxies.  
Even when active, the BHs are faint, rarely reaching their Eddington luminosity ($L_\mathrm{Edd}\sim10^{43}$~erg~s$^{-1}$ for a $\sim10^5 \msun\ $ BH ).  
The first indication of a population of low-mass BHs in late-type galaxies came from the active galactic nucleus (AGN) discovered in NGC~4395 \citep{filippenkosargent1989}, 
which is thought to harbor a $\sim 10^5$~\msun\ BH \citep{filippenkoho2003,petersonetal2005}.  The rediscovery of the AGN in POX 52, also thought to harbor a $\sim 10^5$~\msun\
BH \citep{barthetal2004}, inspired a number of more systematic searches for BHs in this low-mass range.  
Systematic searches through the Sloan Digital Sky Survey (SDSS) have uncovered more than one hundred BHs with $\mbh<10^6$~\msun\ \citep{greeneho2004,greeneho2007a,dongetal2007,dongetal2012}.  
Deeper surveys of late-type spirals with optical and mid-infrared spectroscopy have uncovered a handful of additional examples \citep{shieldsetal2007, satyapal2007,satyapal2009,desrochesho2009,reinesetal2011,reinesetal2012}.  
We are beginning to study the accretion properties \citep{greeneetal2006,greeneho2007a, desrochesetal2009,wrobeletal2008,miniuttietal2010,dongetal2012,ludwigetal2012} and the distribution of host properties for these samples 
\citep{barthetal2005,greeneetal2008, xiaoetal2011,jiangetal2011}.  On the other hand, because of severe bias in the optical selection of these objects, it is prohibitive to calculate a robust space density for them \citep[see review in][]{greene2012}.

Alternate search strategies are needed that circumvent strong optical
bias toward luminous host galaxies and high-Eddington ratio sources.
Techniques using X-rays or mid-infrared spectroscopy have utilized
very small samples to date
\citep{dewangan2008,desrochesetal2009,satyapal2009,gallo2010,leipski2012,ho2012,terashima2012,arayasalvo2012,secrest2012}.
On the other hand, very deep X-ray surveys now exist that reach depths
capable of probing the intermediate-mass regime for BHs.  For
instance, the central area of the Chandra Deep Field-South Survey reaches depths (up to 4 Ms) capable of uncovering $\sim 10^{5-6}
\msun\ $ BHs with $L_{\rm bol}$/$L_{\rm Edd}$ as low as $10^{-3}$ out
to $z \approx 1$ \citep{babic2007}.  So far, most studies of the AGN population in these
deep-survey fields have focused on either the high-redshift luminous population
or those hidden by obscuration. Recently, \cite{xue2012} find that there is a significant fraction of relatively low-mass galaxies that host highly obscured
 AGNs at $z\sim1-3$, which contribute to the unresolved 6-8 keV cosmic X-ray background. These surveys enable
searches for accreting BHs in small stellar mass systems that are not
biased by host-galaxy mass.  Furthermore, such surveys have remarkable
multiwavelength data that greatly facilitate our understanding of the
building blocks of the more massive quasars and their host galaxies.

We use X-ray observations from the Chandra Deep Field-South Survey
and its extended coverage to search for faint AGNs within the
low-mass ($\sim10^{8-9} \msun$) galaxy population.  These are
galaxies most likely to harbor low-mass BHs.  We select candidates
with high X-ray luminosities compared to our expectations based on
their optical luminosities and star formation rates (SFRs).  In this
exploratory work, we highlight three candidates with $0.1 <
z < 0.3$, for which we also have high-quality optical spectroscopy.
From the spectra, we seek additional signs of AGN activity from the
strong line ratios, estimate the level of dust extinction, and place
some limits on the star formation rates.  Using high-resolution
imaging from \emph{HST}/ACS, we present properties of the host
galaxies. A follow-up study (Schramm et al., in preparation) will
present the full population of AGNs associated with low-mass galaxies
up to $z\sim1$.

\section{Data and sample selection}

We first select a sample of low-mass galaxies using the GEMS survey
\citep{rix2004} in the Extended Chandra Deep Field-South (E-CDF-S)
field that provides HST/ACS imaging over 0.3 deg$^2$.  Photometric redshifts are provided by the MUSYC survey (Cardamone et
al. 2010) that are based on 32 bands of optical and infrared imaging.
Due to the wealth of spectroscopic programs in the E-CDF-S \citep[e.g.][]{szokoly2004,vanzella2008,popesso2009,treister2009,silverman2010,cooper2012}, we
are able to replace photometric redshifts with spectroscopic values if
available.  We start by selecting galaxies with $z < 1$ such that one
of the two HST filters falls above the 4000 \AA\, break for
sensitivity to older stellar populations.  Stellar masses are
determined from the galaxy absolute magnitude $M_V$ and rest-frame
color $B-V$ as measured from HST imaging, following the prescription
of \citet{bell2003}.  Our stellar mass estimates match well with those from the MUSYC survey \citep{cardamone2010} as derived from broad-band spectral energy distribution (SED) fitting, for galaxies with secure spectroscopic redshifts that are in agreement with the photometric redshifts.

\begin{figure}
\begin{center}
\includegraphics[angle=0,width=120mm]{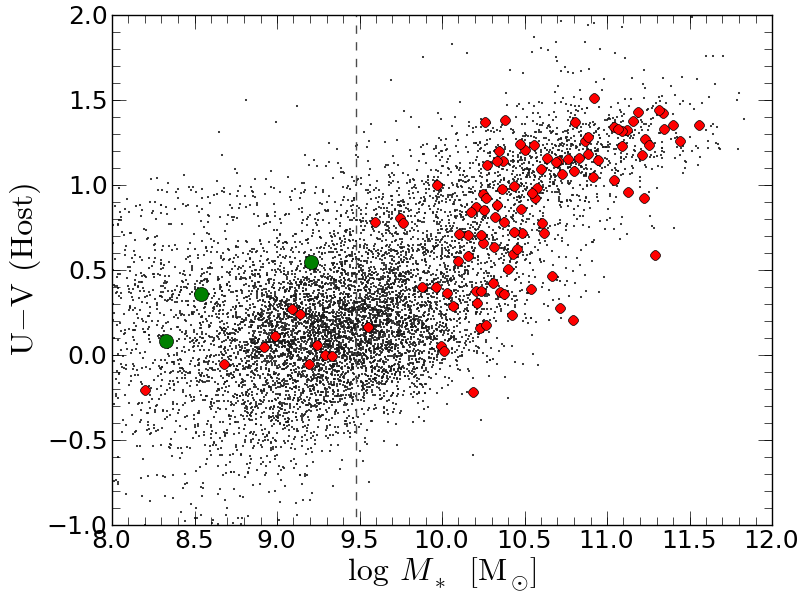}
\end{center}
\caption{Rest-frame U-V as a function of stellar mass for galaxies ($z<1$) in the GEMS survey.  All X-ray sources with $L_X<10^{44}$ erg s$^{-1}$ and $z_\mathrm{spec}<1$ are shown in red while those of interest for this study at $z<0.3$ are marked in green.
The dashed line indicates our selection criterion of $M_*\ <3\times10^9 \msun$.  }
\label{fig:cm}
\end{figure}

In this sample, there are $\sim5200$ galaxies with $M_*\ <3\times10^9
\msun$.  The galaxy cut-off mass we choose is somewhat arbitrary, but
does focus on a stellar-mass regime where only a few AGNs are already known [see, e.g., Figure 1 of \cite{xue2010}; note the $3\times10^9\,\msun$ cutoff line in their Figure 1].
Specifically, low-mass black holes found through optical selection
using the SDSS \citep{greeneho2007b} are typically
found in higher mass galaxies with stellar masses of a few times
$10^9-10^{10}$ $\msun$ \citep[see also][]{dongetal2012}. The stellar-mass regime below $\sim10^9$ $\msun$ has yielded essentially no AGNs
in optically-selected searches \citep{ho1997,kauffmann2003,barth2008}.
Also, based on a naive extrapolation of the BH-to-bulge mass
relation of \citet{haeringrix2004}, these galaxies are likely to
harbor a black hole with $M_{BH}<2\times10^6$ $\msun$.  Finally, the
upper limit on the stellar mass of the parent galaxy population
focuses our search for low-mass black holes at the low-mass end of
the blue cloud, a region of parameter space in the color-mass plane
(see Figure~\ref{fig:cm}) where galaxies are likely to be forming stars rapidly.

Multi-wavelength catalogs of X-ray sources in the CDF-S \citep{luo2010,silverman2010,xue2011} provide robust matches to optical counterparts, 
including low-mass galaxies, based upon a likelihood-ratio matching routine.  We find 27 galaxies with $z <
1$, $M_* < 3 \times 10^9$~\msun, and an X-ray detection. We note that all X-ray sources have a wavdetect
false-positive probability of $10^{-8}$. Of these 27 galaxies, we
concentrate here on the 3 galaxies (out of a parent population of $\sim2100$) with $z < 0.3$ and existing
optical spectra and a high ratio of X-ray to
optical luminosity.  At these redshifts our spectra cover four strong
optical emission lines: H$\beta$, [O {\small III}]~$\lambda \lambda
4959, 5007$, H$\alpha$, and [N{\small II}]~$\lambda \lambda 6548, 6584$,
which will be used for further classification and analysis.

\begin{figure}
\includegraphics[angle=0,width=50mm]{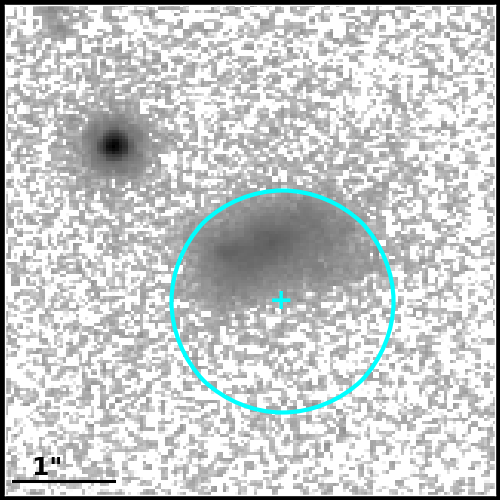}
\includegraphics[angle=0,width=50mm]{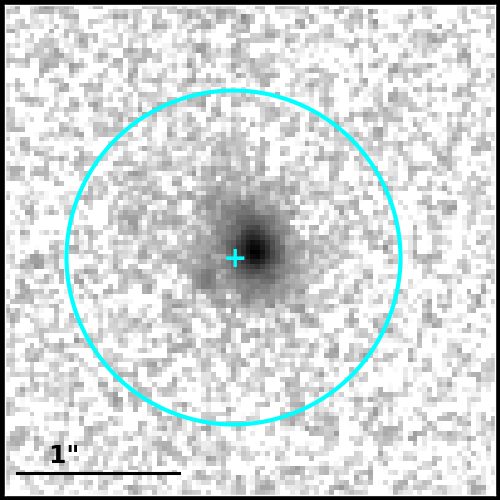}
\includegraphics[angle=0,width=50mm]{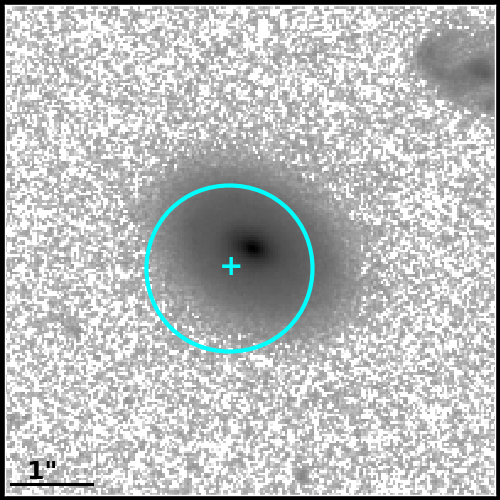}
\caption{HST/ACS images of three low-mass galaxies determined to be the optical counterparts of the $Chandra$ detections (left: 231, middle: 476, right: 312).  
A blue cross marks the centroid position of the X-ray detections.  A circle indicates an effective search radius of 1$\arcsec$.}
\label{fig:optical} 
\end{figure}

\begin{figure}
\begin{center}
\includegraphics[angle=0,width=10cm]{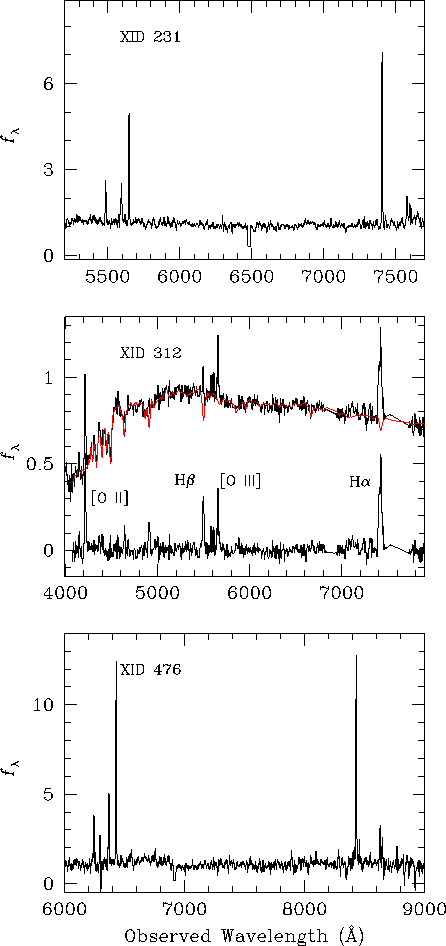}
\end{center}
\caption{Optical spectra taken with either Keck/DEIMOS (XID=231 and 476) or VLT/FORS2 (XID=312). Units are given in relative flux. For XID-312 we plot the spectrum before
and after stellar continuum subtraction. The best-fit continuum is shown in red.}
\label{fig:spectra} 
\end{figure}

In this paper we present three candidate AGNs in low-mass galaxies, selected to have a high ratio of X-ray to
optical luminosity as detailed below. 
If confirmed, these galaxies would be among the lowest-mass host galaxies known \citep[see
also][]{barth2008,reinesetal2011}. Two of them (XID-476, XID-231) are detected in the shallower
E-CDF-S coverage \citep{lehmer2005} and the third (XID-312) within the central 4
Ms area \citep{xue2011} with more than 100 counts in the full band (0.5-8 keV). In Figure~\ref{fig:optical}, we show postage-stamp HST/ACS images of the three low-mass galaxies while indicating the X-ray 
centroid and an effective search radius for an optical counterpart that demonstrates the likelihood of these associations.  In the remainder of this paper, we combine the \emph{HST} imaging 
(Figure~\ref{fig:optical}), the optical spectroscopy from 
Keck/DEIMOS \citep{silverman2010} and VLT/VIMOS \citep{szokoly2004}
(Figure~\ref{fig:spectra}), and the broad-band SED to argue that each of these sources 
is a viable candidate to host a low-mass AGN (\S 3).  



 \section{Disentangling X-ray emission from AGNs and normal galaxies}

 We present our case that these three X-ray detected
 galaxies are most naturally explained as being powered by accretion onto a massive black
 hole.  There are many other processes that lead to X-ray emission
 from galaxies, including an unresolved contribution of low-mass and
 high-mass X-ray binaries or thermal emission from hot gas
 \citep[e.g.,][]{fabbiano2006}.  In particular, the high-mass X-ray
 binary population is likely to contribute significantly to those
 galaxies with high SFRs.  Fortunately, there
 are many studies
 \citep[e.g.,][]{ranalli2003,osullivan2003,lehmer2010,boroson2011} of X-ray
 emission from inactive galaxies (with no AGN) that enable us to
 determine whether the level of X-ray emission is characteristically
 higher than that expected for galaxies of a given luminosity (i.e.,
 stellar mass) and SFR. 

\begin{figure}
\begin{center}
\includegraphics[angle=0,width=120mm]{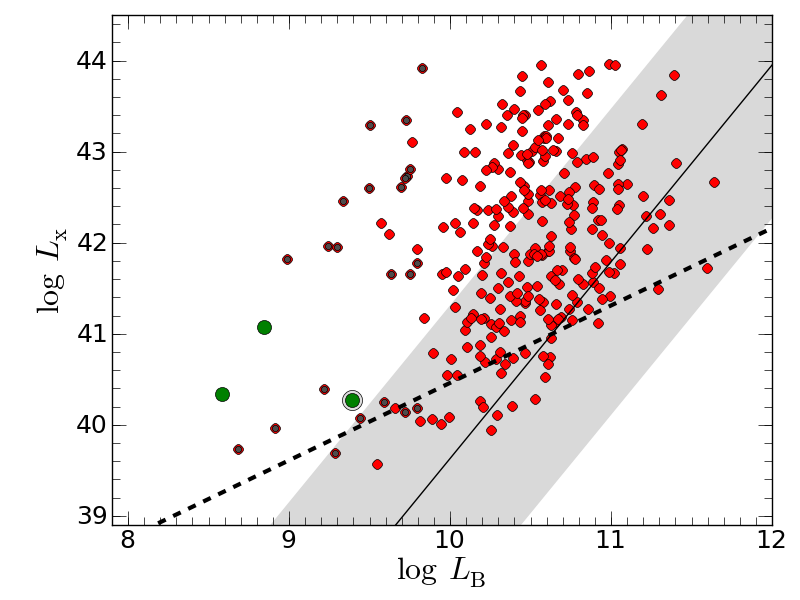}
\end{center}
\caption{X-ray luminosity versus B-band luminosity for X-ray sources in the E-CDF-S \citep{lehmer2005} covered by HST with $L_X<10^{44}$ erg s$^{-1}$ and $z_\mathrm{spec}<1.0$.  Three X-ray sources associated with galaxies having $z<0.3$ are shown in green.
Sources with stellar masses below $M_*<3\times10^{9}$ M$_{\sun}$ have an additional gray circles. The solid line indicates where normal elliptical galaxies /star-forming typically lie (O'Sullivan et al. 2003). The dashed line shows the 1$\sigma$ upper bound for star-forming galaxies from Lehmer et al. 2010. We have encircled the position of
XID-312 as the variable source showing a broad H$\alpha$ line. }

\label{fig:select}
\end{figure}

 We first show that our candidates have excess X-ray emission compared
 to that expected from early-type galaxies based on their optical
 continuum luminosities and then we show that the sources have even more anomalous 
X-ray luminosity given their inferred SFRs.  In Figure~\ref{fig:select}, we
 plot the distribution of X-ray luminosity as a function of $B$-band
 luminosity.  The $1 \sigma$ region of parameter space where X-ray emission is
 characteristic of early-type galaxies \citep{osullivan2003} is
 shaded. For reference, we also include all 
galaxies with X-ray detections ($L_X<10^{44}$
 erg s$^{-1}$) and $z<1$ in red. The dashed line gives the $1\sigma$ upper boundary for star-forming galaxies (Lehmer et al. 2010). 
 We show the location of our three low-mass galaxies at $z<0.3$ that are
 X-ray emitters (green points). Although XID-312 is just above our selection criteria in the
 $L_X-L_B$ plane, the galaxy has been reported to be X-ray variable in
the recent study by \citet{young2012}, who concluded that the observed
variability cannot be due to binary populations or ultraluminous
sources but rather is best explained by an accreting black hole. Its X-ray hardness ratio, defined as HR=(H-S)/(H+S), where S and H are the soft (0.5-2 keV) and hard (2-8 keV) band net counts
is HR=-0.2 \citep{xue2011}. The observed hardness ratio suggests some absorption by neutral gas with a density of about $N_H\sim10^{22} \mathrm{cm^{-2}}$ \citep[see Figure 9 in][]{silverman2005}.
 There are 22 low-mass ($ M_*
 <3\times10^9$ $\msun$) galaxies at $z<1$ that have an $L_X$ to $L_B$ ratio
 more than $1\sigma$ away from the typical relation for early-type
 galaxies, including the three galaxies at $z<0.3$ that are the
 focus of this study. 

Now we turn to the optical line emission, (see Figure \ref{fig:spectra}) for additional clues as to
the nature of these sources.  We use traditional emission-line ratio
diagnostics (e.g., Kewley et al. 2006) to determine the dominant source of
photo-ionization in these galaxies (AGN activity or star formation).
We perform a line-fitting routine to measure the strengths of the
emission lines. We first fit the spectrum with a
low-order polynomial to model the continuum adjacent to the emission
line and then subtract it from the spectrum.  In the cases of XID-476 and XID-231, we do not detect any 
starlight in the continuum.   Line fluxes are
measured by integrating our best fit obtained for a multi-Gaussian
model.  For XID-312, there is significant galaxy continuum, and so we use 
GANDALF \citep{sarzietal2006} with BC03 models \citep{bruzual2003} to model and subtract the galaxy continuum 
before applying the same line-fitting routines.
In case of XID-476 and XID-231, a single Gaussian is
sufficient to model the line profiles, but for XID-312 with its broad
component in H$\alpha$ and line blending with [N {\small II}], a multi-component
fit is necessary.  We scale the spectra to the flux in the broad ACS F606W filter.

\begin{figure}
\begin{center}
 \includegraphics[angle=0,width=15cm]{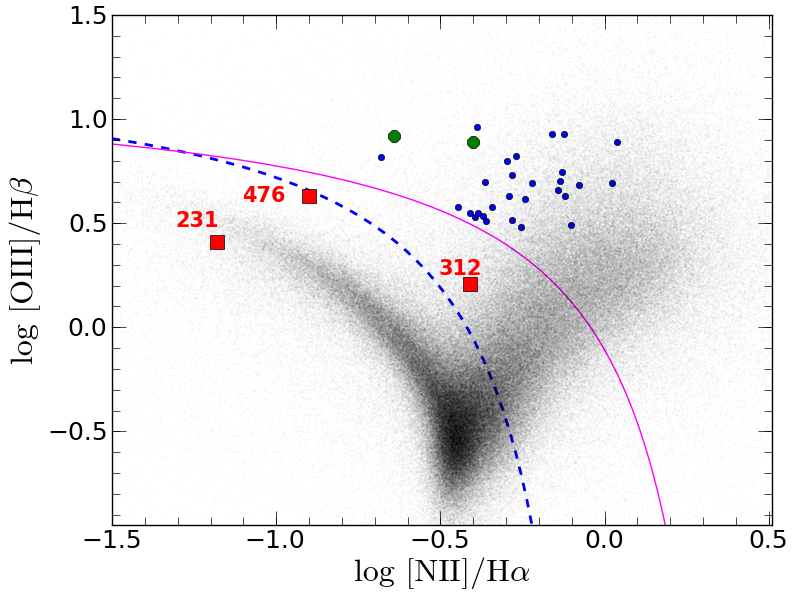}
\end{center}
\caption{Line-ratio diagram for [O III]5007/H$\alpha$ versus [N II]6583/H$\alpha$. NGC 4395 and POX 52 are shown as green circles. The blue circles mark the positions
of low-mass Seyfert II galaxies from Barth et al. (2008). 
The solid curve is the 'maximum starburst' line from Kewley et al. (2006), and the dashed line is from Kauffmann et al. (2003).  }
\label{fig:BPT} 
\end{figure}

In Figure \ref{fig:BPT}, we plot the line ratios for [O {\small III}]/H$\beta$
versus [N {\small II}]/H$\alpha$.  We find that the line ratios are not typical
of Seyfert 2 galaxies; all fall below the demarcation line of Kewley
et al. (2006) that cleanly separates star-forming and AGN dominated
galaxies.  Two of the three galaxies are close to the demarcation line
of \citet{kauffmann2003} that was established to include galaxies
having a composite nature (AGN + star formation).  For comparison, the
locations of low-mass Seyfert 2 galaxies, as described in Barth et
al. (2008), are indicated by small blue circles; these galaxies have
been selected to be bona-fide AGNs by their emission-line ratios.  Two
of our three galaxies are located at low [NII]/H$\alpha$ and high
[OIII]/H$\beta$.  Given the low stellar masses of our sample, these low 
ratios are most naturally explained as a metallicity effect \citep{tremonti2004,groves2006,ludwigetal2012,stern2012}.
From the [N{\small II}]/H$\alpha$ ratio, using the empirical
calibration of Nagao et al. (2005), we estimate the metallicities of the
three host galaxies to be in the range of 0.3-0.6 solar metallicity.
From the total emission-line intensity ratios I(H$\alpha$)/I(H$\beta$), 
we estimate that all three galaxies show internal extinction ranging from A$_V$=0.9-1.6 mag. 

In the case of XID-312, we report here a possible identification
of a broad H$\alpha$ line in the low-resolution optical spectrum from
Szokoly et al. (2004). We performed the decomposition of the H$\alpha$
profile into a Gaussian for the broad component and the [O {\small III}] 
line profile to model the narrow H$\alpha$ and [N {\small II}]. We
estimate that the broad component has a FWHM of $\sim1400$ km~s$^{-1}$ (corrected for instrumental resolution)
but caution that the resolution of this spectrum is low, and 
the continuum is dominated by galaxy light.
The corresponding black hole mass would be log $M_{BH/M_\odot}$=5.3 using the
calibration of \citet{greeneho2005}. Although X-ray variability of this source and the broad H$\alpha$ line are strong
indications of an AGN, the overall line ratios place this galaxy in the composite (SF + AGN) region in the BPT diagram
\citep{stern2012} (see Figure \ref{fig:BPT}).

\begin{figure}
\begin{center}
\includegraphics[angle=0,width=8.5cm]{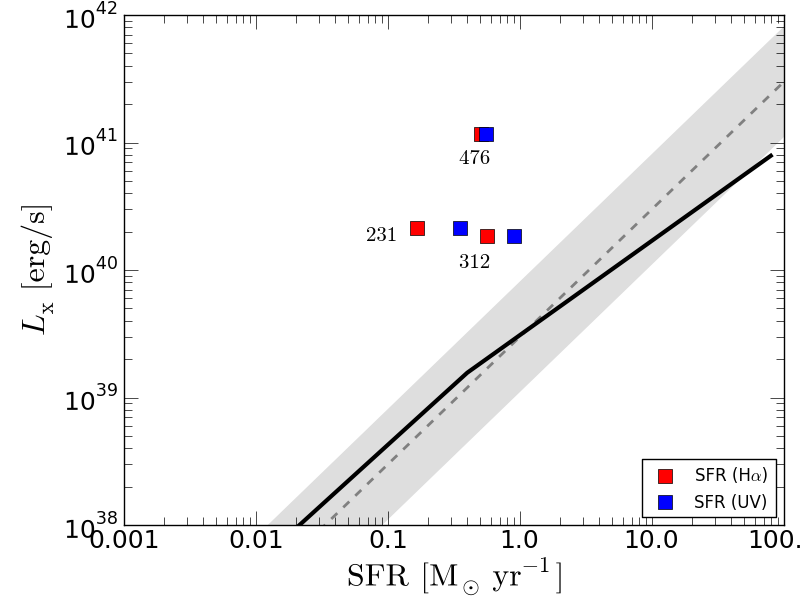}
\end{center}
\caption{X-ray luminosity versus star-formation rate estimated from H$\alpha$ luminosity (red) and UV continuum measured from the GALEX-NUV photometry. 
We make the assumption that all measured flux is due to star-formation and no contribution is 
due to the presence of an AGN. This includes the possible contamination of scattered light to the measured UV continuum flux. The dashed line shows the best fit for 
local galaxies by Mineo et al. (2011) with an scatter of 0.4 dex marked as the shaded area. The solid line shows the relation from \citet{lehmer2010}.}
\label{fig:XrayEL_sfr} 
\end{figure}

We can use the narrow line emission as a further test of the origin 
of the X-ray emission. We take the total H$\alpha$ luminosity as an 
upper limit to the SFR, under the conservative assumption
that there is no contribution to the line emission from an AGN.  SFRs
are corrected for Galactic and internal extinction based on the
measured Balmer decrement assuming a \citet{calzetti2000} extinction
curve.  Using H$\alpha$ luminosity, we find that the upper limits to the SFRs are 
$\sim$0.2-0.6 $M_\odot$~yr$^{-1}$, and are consistent with those seen in galaxies with
masses of $\sim10^9$ $M_\odot$ \citep{elbaz2007}. As a second SFR
estimator, we use the extinction corrected GALEX-NUV flux measurement, 
which gives a consistent SFR estimation again assuming no contribution from 
accretion.  We find that for these upper limits on the SFRs, the X-ray
emission is $>~4\sigma$ (XID-231, XID-476) and $\sim3\sigma$ (XID-312) larger than expected for
star-forming galaxies (see Figure \ref{fig:XrayEL_sfr}) using the relations given in \citet{mineo2011}
and \citet{lehmer2010}.

Therefore, there are two possible explanations for our sources.
First, they have one or several highly super-Eddington X-ray binaries,
or ultra-luminous X-ray sources.  Second, they contain
accreting nuclear black holes.  In either case, they are interesting outliers from the $SFR-L_\mathrm{X}$ relations.  
In the case of XID-312, the X-ray variability and broad H$\alpha$ line both suggest that the source is powered by a massive BH.  
We test the possibility of some super-Eddington X-ray binaries through a simple Monte-Carlo test. We find that the probability for the two remaining galaxies to be outliers ($>4\sigma$) in the $L_\mathrm{X}$-SFR relation is $<0.2\%$ 
based on our parent population of $\sim 2100$ galaxies at $z<0.3$. We expect at most one false positive detection at $4\sigma$ within our parent. 
For this test we make the conservative assumption that we have a uniform X-ray sensitivity of the 4 Ms survey for the full parent population.

\section{Host-galaxy morphology}

\begin{figure}
\begin{center}
\includegraphics[angle=0,width=8.5cm]{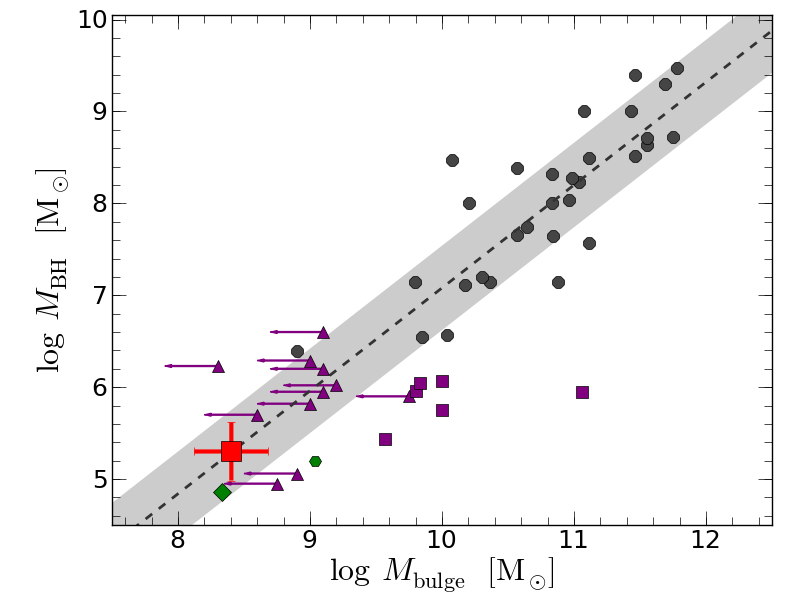}
\end{center}
\caption{\mbh\,as a function of bulge mass. The estimate for XID-312 is shown as a red solid square. We overplot the local inactive sample from \cite{haeringrix2004} (gray symbols) as well as their best fit (dashed line) and a typical 0.3 dex uncertainty (shaded area). 
The low-mass BH sample from \cite{greeneetal2008} is shown as purple symbols. The solid squares mark compact smooth objects (likely to be ellipticals or spheroidals) and the filled triangles indicate the bulge components of 
disk galaxies. We use the smaller sample from \cite{greeneetal2008} rather than the \cite{jiangetal2011b} 
sample because Greene et al.\ convert to stellar mass in a way that is more directly comparable 
with our measurements. We add the bulge mass estimate for POX 52 (green hexagon) and the total mass estimate for NGC 4395 (green diamond)}
\label{fig:MBH-Mbulge} 
\end{figure}

We examine the host morphologies of each AGN candidate  to provide some clues as to the formation mechanism of 
these low-mass black holes and their hosts. XID-312 falls within the GOODS-South area with coverage with five \emph{HST}
filter bands (ACS:F435W, F606W, F775W, F850LP, and F814W). 
We could not find a point source in a two-dimensional fit to the \emph{HST}
images (given a limiting host-to-nuclear flux ratio Host/Nuclear$>$50) that included an empirical point spread function model (PSF) created from nearby stars.  
We use a composition of three S\'{e}rsic profiles to fit the galaxy of XID-312. We
use an n=1 and n=4 model to describe the disk and the bulge of the
galaxy and an extended component with n=0.42. We estimate the bulge to
total ratio B/T=0.16 (in the $z-$band) showing that the galaxy is mostly
disk dominated. Although the galaxy is disk dominated, we do not see
any signs of spiral arms, even at the resolution of \emph{HST}. For the 
estimated black hole mass of $2\times 10^5$ $M_\odot$ from the broad H$\alpha$ emission line and the stellar bulge mass, we find for the 
measured B/T, XID-312 falls 
onto the black hole mass - bulge mass relation established for local
galaxies by \citet{haeringrix2004} (see Figure\,\ref{fig:MBH-Mbulge}). The bulge component of XID-312 appear to be consistent with the relation at higher mass
as well as the bulge components of disks from \cite{greeneetal2008} with similar BH masses.

The morphological analysis of XID-476 using GALFIT to fit a S\'{e}rsic model shows that its host galaxy is an
early type galaxy based on the best-fit S\'{e}rsic index ($n_{\mathrm{Sersic,F850LP}}=3.3$).
We do not find a point source (Host/Nuclear$>$40) in the
center of the galaxy.  Since we do not have a broad emission line to use to estimate the
BH mass, we convert the stellar bulge mass into a
BH mass using the relation from \citet{haeringrix2004}, yielding
a BH mass of $2.5\times 10^5$ $M_\odot$.

XID-231 has the lowest stellar mass of all three candidates (log
M$_{*}=8.3$) and shows the most peculiar host among all three
galaxies. We find no sign of a point source (Host/Nuclear$>$40) from our two-dimensional
imaging analysis. Although the best-fit single S\'{e}rsic index is n=2.4,
the residuals are significant and suggest that a single S\'{e}rsic model
is insufficient to describe the host morphology. Adding further S\'{e}rsic
models to the fit reduces the residuals but we cannot identify
a prominent bulge or disk component even by fixing some of the
parameters.  We conclude that this galaxy is of an irregular type. We estimate a
black hole mass of $1.5\times 10^5$ $M_\odot$ based on the total 
estimated stellar mass, given the morphological irregularity.
The S\'{e}rsic indices together with the stellar masses and X-ray properties of the galaxies are listed in Table \ref{sample}.

\begin{figure}
\begin{center}
\includegraphics[angle=0,width=5.3cm]{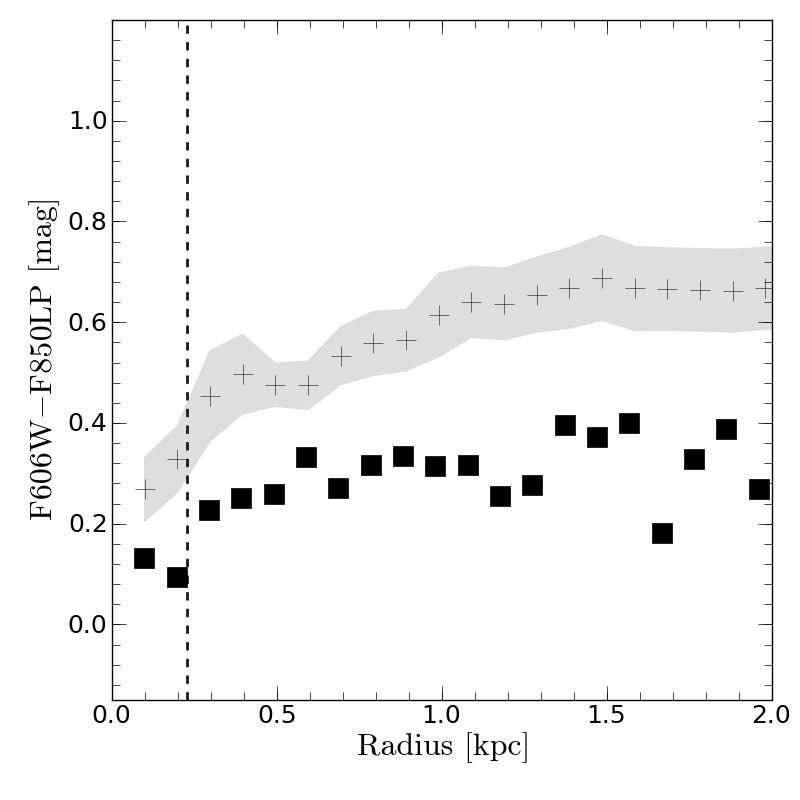}
\includegraphics[angle=0,width=5.3cm]{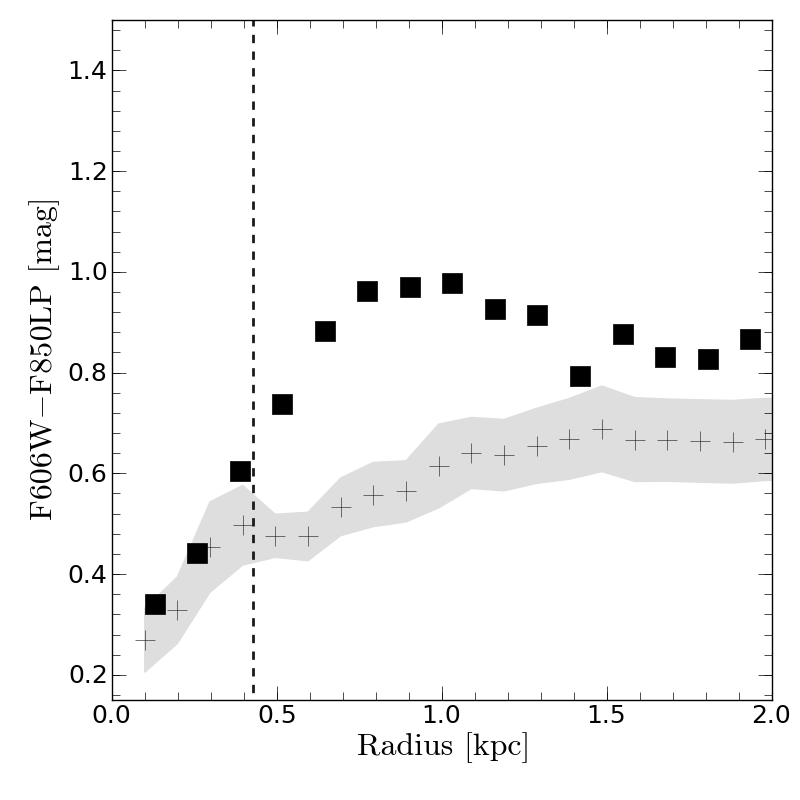}
\includegraphics[angle=0,width=5.3cm]{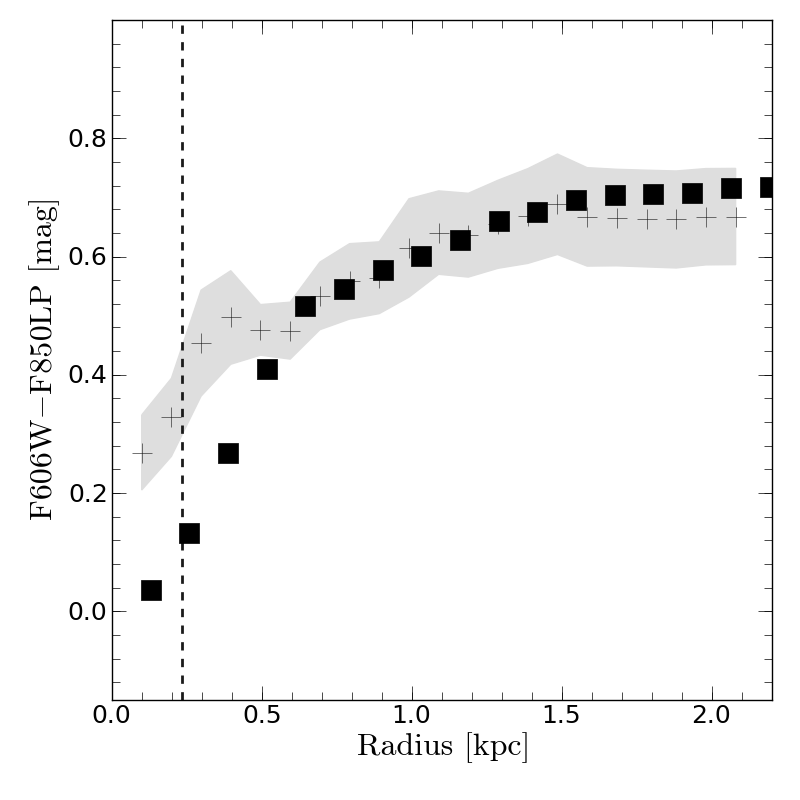}
\end{center}
\caption{Observed F606W-F850LP color profile of (from left to right) XID-231, XID-476 and XID-312 (solid squares). For comparison we show the average color profile of a redshift and mass matched
sample of inactive galaxies (crosses) and the error of the mean as the gray shaded area. The vertical line shows the typical size of the PSF.}
\label{fig:HST_colorprof} 
\end{figure}

Finally, we make use of the different HST filter bands and look at the radial color profiles of the host galaxies.
The analysis of XID-312 reveals a blue core with a F606W-F850LP color gradient of 0.5 mag (see Figure \ref{fig:HST_colorprof}). The F435W-F606W color gradient
is even steeper (1.25 mag) between the center and a radius of $\sim$0.7 kpc.   
It may be that the blue light is scattered from an AGN \citep[see, e.g.,][]{zakamska2005}. In the case of XID-476, we find a similar blue core as for XID-312. The F606W-F850LP color
shows a steep gradient of 0.7 mag toward the center starting
at a similar radius of about 0.7 kpc (see Figure \ref{fig:HST_colorprof}). Although, the
F606W-F850LP color gradient of XID-231 shows a slightly bluer color (0.25 mag) (see Figure \ref{fig:HST_colorprof})
toward the center, the overall color profile remains flat.
We compare the observed color gradients of our candidates with an average color profile of a redshift and mass matched sample ($\sim10^9$ $M_\odot$) of inactive galaxies 
shown as a gray shaded profile in Figure \ref{fig:HST_colorprof}. At $r < 0.7$ kpc the color gradient of XID-312 and XID-476 is much steeper than the gradient of the comparison sample 
resulting in a color difference of $\sim0.3$ mag. We interpret this as evidence for a contribution from the AGN. 
The color gradients of these two sources also differ from the color profiles observed in point-source free unobscured X-ray selected AGN host galaxies at $0.5<z<1.5$ \citep[][]{ammons2011}.
This might reflect the complexity in disentangling AGN and SF activity on small scales. The overall profile of XID-231 is comparable with the average profile of inactive galaxies 
although the color is overall bluer.

\section{Conclusions}

We have identified three galaxies at $z<0.3$ with stellar masses below
$3\times 10^9 M_\odot$ that likely harbor low-mass BHs.  
If confirmed, these would be among the lowest-mass galaxies known
to host an AGN. 
They are roughly an order of magnitude more luminous in the X-rays than expected for highly super-Eddington X-ray binaries, or ultra-luminous X-ray sources 
based on their continuum luminosities or their estimated SFRs.  
Assuming the empirical relation between BH mass and
stellar bulge mass \citep[e.g.,][]{haeringrix2004}, we can estimate the BH masses to be of order $\sim2\times10^5
M_\odot$. 

Accreting BHs in low-mass galaxies may contribute significantly to the hard X-ray background 
\citep{xue2012} so studying the host galaxy properties in detail is an important task. 
The morphological analysis of the \emph{HST} imaging data shows that
each host is different. While XID-476 has an early type host galaxy,
XID-312 shows an intermediate and XID-231 an irregular host
galaxy. Although all three galaxies show different morphologies, XID-312 and XID-476 have
in common a blue core in the center of the galaxy ($<0.7$ kpc). This
suggests that the ionizing source is located in the center of the
galaxy and could therefore be an AGN. The diversity in the host morphology provides an insight into how BHs interacted with their host environment in the early 
universe \citep{reinesetal2011,jeon2012}.


Of the three low-mass AGN candidates, XID-312 shows the strongest evidence of having an accreting BH.
Besides the excess in the X-ray emission (compared to SFR and optical continuum luminosity) and the X-ray variability \citep{young2012}, this object shows an indication of a broad H$\alpha$ emission line
with a FWHM$=1400$ km~s$^{-1}$ resulting in a corresponding BH mass of $\sim2\times 10^5$ $M_\odot$. 
This estimate is consistent with the BH mass prediction using the local BH mass - bulge mass. Although only a single object with independent BH and bulge mass estimate, XID-312 nicely extends the local relation at the low-mass end.  
For the other two sources the strong observed X-ray emission (although based on typically $<10$ X-ray counts), more than $4\sigma$ above the expected emission from star-forming galaxies,
is the strongest argument for hosting an AGN. Performing a conservative Monte Carlo test, we find that the probability for just being outliers as star-forming galaxies is $<0.2\%$ based on our parent population of $\sim 2100$ galaxies at $z<0.3$. More definitive proof that these two galaxies harbor AGNs could come from additional epochs of X-ray observations looking for variability similar to XID-312. 
In addition resolved spectroscopy of the core region of the host looking for notable line ratios might also give further evidence of the presence of an AGN.

Eventually we want to understand whether all low-mass galaxies, such as those studied in 
this paper, host supermassive BHs.  However, the story is complicated because as one goes to 
lower stellar mass, both the total occupation fraction (number of galaxies hosting a black hole; 
\citep[e.g.,][]{volonterinatarajan2009,bellovary2010} and 
the distribution of Eddington ratios \citep[e.g.,][]{schulzewisotzki2010,nobuta2012,aird2012} may change in principle. 
Without independent constraints on the Eddington ratio distribution, we can derive only 
crude limits on the occupation fraction \citep[e.g.,][]{greene2012}.  However, we 
hope to gain more insights, and better statistics, from studying the full sample of 27 potential 
low-mass BHs residing low-mass galaxies identified in this work.

 

\acknowledgments
The authors fully appreciate the useful discussions with Aaron Barth and Masayuki Tanaka that improved the paper.
This work was supported by the World Premier International Research
Center Initiative (WPI Initiative), MEXT, Japan. YQX acknowledges the financial support of the Thousand Young Talents (QingNianQianRen) program (KJ2030220004), the USTC star
tup funding (ZC9850290195), and the National Natural Science Foundation of China through NSFC-11243008. W. N. Brandt and B. Luo acknowledge the NASA ADP grant NNX10AC99G and the CXC grant AR3-14015X.

\clearpage



\begin{deluxetable}{llllllllllll}
\tabletypesize{\scriptsize}
\tablecaption{AGNs in low-mass galaxies \label{sample}}
\tablewidth{0pt}
\tablehead{
\colhead{ID}&\colhead{RA}&\colhead{DEC}&\colhead{$Redshift$}& \colhead{$R_{AB}$}&\colhead{log $M_*$\tablenotemark{a}}&\colhead{$n_{Sersic}$}&\colhead{X-ray counts\tablenotemark{b}}&$log~f_X$\tablenotemark{b}&\colhead{$log~L_{X}$\tablenotemark{b}}&\colhead{$log~M_{BH}$}}
\startdata
231\tablenotemark{c}&52.98483&-27.736667&0.128&22.7&8.3 (7.7)&2.4&9.4$^{+4.8}_{-3.4}$&-15.3&40.3&5.2\\
476\tablenotemark{c}&53.207625&	-28.006750&	0.285&	23.6&8.5 (8.2)&3.3&8.8$^{+5.0}_{-3.7}$&-15.3&41.1&5.4\\
312\tablenotemark{d}&53.094917	&-27.873389	&0.131&	20.4&9.2 (8.8)&2.9&$112.4^{+15.1}_{-13.9}$&-15.4&40.3&5.3 (virial)\\
\enddata
\tablenotetext{a}{Mass estimate based on GEMS \citep{rix2004} while that in parenthesis is from the MUSYC survey \citep{cardamone2010}.}
\tablenotetext{b}{0.5-8.0 keV band; rest-frame $L_X$}
\tablenotetext{c}{X-ray ID given in \cite{lehmer2005}}
\tablenotetext{d}{X-ray ID given in \cite{xue2011}; XID=521 in \citep{giacconi2002,szokoly2004}}
\end{deluxetable}
\end{document}